\documentclass{eas}
\usepackage{graphicx}
\usepackage{floatrow}
\begin{document}

\title{Interstellar shock studies:\\ the SOFIA/GREAT contribution} 
\author{Antoine Gusdorf}\address{LERMA, Observatoire de Paris, PSL Research University, CNRS, UMR 8112, F-75014, Paris, France; Sorbonne Universit\'es, UPMC Univ. Paris 6, UMR 8112, LERMA, F-75005, Paris, France}
%
%
\begin{abstract} 
Shocks are ubiquitous in the interstellar medium of galaxies, where they contribute to the energetic balance and to the cycle of matter, and where they are thought to be the primary sites for cosmic rays acceleration. Most of the time: in jets and outflows, supernova remnants, or colliding flows, they are linked with star formation. The study of shocks is hence a powerful tool to probe the evolution of the interstellar medium and to better understand star formation. To these aims, the most precise observations must be compared with the most precise models of shocks. The SOFIA/GREAT instrument represents a powerful observational tool to support our progresses, as it allows to observe numerous shock tracers in the far-infrared range.
\end{abstract}
\maketitle
\section{Introduction}

The interstellar medium (ISM) is constantly evolving due to unremitting injection of energy in various forms. Energetic radiation transfers energy to the ISM: from the UV photons, emitted by the massive stars, to X- and $\gamma$-ray ones. Cosmic rays are another source of energy. Finally, mechanical energy is injected through shocks or turbulence. Shocks are ubiquitous in the interstellar medium of galaxies. They are associated to star formation (through jets and bipolar outflows), life (via stellar winds), and death (in AGB stellar winds or supernovae explosion). The dynamical processes leading to the formation of molecular clouds also generate shocks where flows of interstellar matter collide. Because of their ubiquity, the study of interstellar shocks is also a useful probe to the other mechanisms of energy injection in the ISM. This study must be conducted in order to understand the evolution of the interstellar medium as a whole, and to address various questions detailed in this review. To this aim, it is paramount to interpret the most precise observations with the most precise shock models. From the observational point of view, the SOFIA telescope, and particularly the GREAT receiver and the spectral resolution it offers, represents a powerful tool to better address the above questions, as it allows to observe numerous shock tracers in the far-infrared range.

\section{Low-mass star formation}
\label{sec:lmsf}

\begin{figure}
\floatbox[{\capbeside\thisfloatsetup{capbesideposition={left,top},capbesidewidth=0.4\textwidth}}]{figure}[\FBwidth]
{\caption{Low- (left-hand panels, BHR71, Gusdorf {\em et al.\/} \cite{Gusdorf151}) and intermediate-mass (right-hand panels, Cep E, Lefloch {\em et al.\/} \cite{Lefloch15}) star formation: maps (upper panels) and $^{12}$CO spectra (lower panels) showing the SOFIA/GREAT observations. The BHR71 image is an overlay of 8~$\mu$m emission (colours, \textit{Spitzer}/IRS) with CO (3--2) (white contours, APEX). The Cep E image from PdBI shows the jet, the terminal bowshock, and cavity in the southern lobe. Each spatial element corresponds to a spectral range centered on the vertical dashed lines of the lower panel, in the same colours.}\label{figure1}}
{\includegraphics[width=0.6\textwidth]{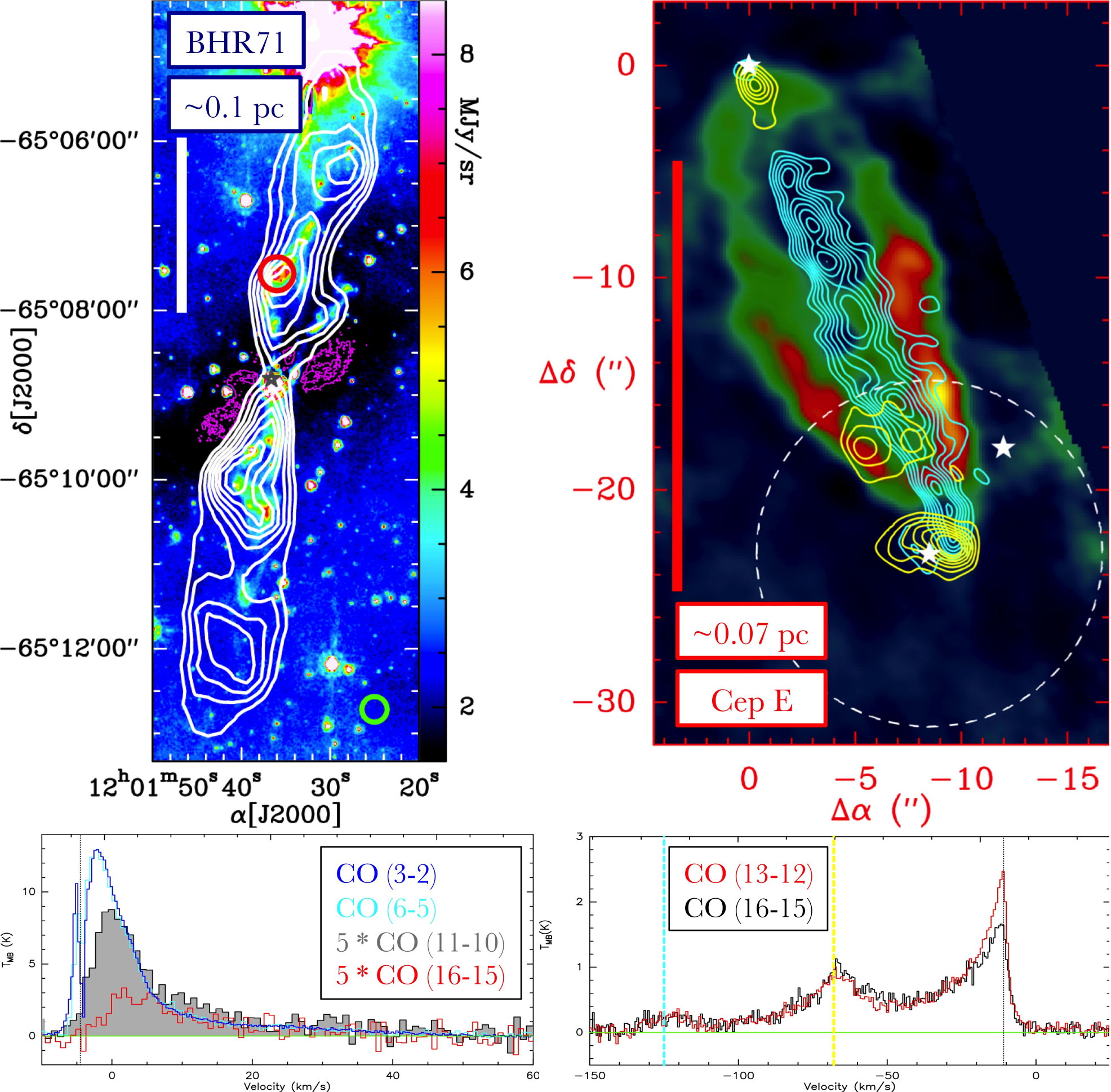}}
\end{figure}

The paradigm for the formation of low-mass stars has been established over the past decades. It involves an accretion of matter through a disk-like structure, as well as ejection through jets and outflows, generally bipolar (e.g. Fig.~\ref{figure1}). However, a lot of shock-related questions remain: what are the ejection mechanisms of the molecular gas ? what are the mechanisms leading to the peculiar chemistry observed in numerous low-mass star forming environments ? What are the energetic impacts of jets and outflows on their environment ? Can cosmic rays be accelerated in the associated shocks ? To address these questions, the observation of emission from high-$J$ rotational lines of CO is an important probe to the temperature and density conditions. Indeed, CO is an abundant molecule whose levels for which $J_{\rm up} > 7$ trace the warm ($E_{\rm up} > 155$~K) and typically shocked medium. Today, SOFIA/GREAT is the only telescope that allows for their observation. The chemically rich outflows L1157 (Eisl\"offel {\em et al.\/} \cite{Eisloeffel12}) and BHR71 (Gusdorf {\em et al.\/} \cite{Gusdorf151}) have been observed with SOFIA/GREAT. In the latter case, their comparison with state-of-the-art shock models led to precise determinations of masses, momentum and energy transfers. Combined with H$_2$ and SiO observations, they also allowed to better constrain the silicon chemistry.

\section{High- and intermediate-mass star formation}
\label{sec:hmsf}

\begin{figure}
\floatbox[{\capbeside\thisfloatsetup{capbesideposition={left,top},capbesidewidth=0.5\textwidth}}]{figure}[\FBwidth]
{\caption{Massive star formation: Cepheus A (left-hand panels) and G5.89--0.39 (right-hand panels). For Cepheus A, the 2.12~$\mu$m map from Cunningham{\em et al.\/} \cite{Cunningham09} shows the complex outflow system, while the spectra show the OH observations by SOFIA/GREAT at 117 and 163~$\mu$m. For G5.89--0.39, the large-scale (top, Hunter {\em et al.\/} \cite{Hunter08}) and zoomed-in (middle, Su {\em et al.\/} \cite{Su12}) maps also illustrate the complexity of the star forming region, while the spectrum shows the velocity-resolved emission/absorption structure in the [OI] 63~$\mu$m line, as seen by SOFIA/GREAT.}\label{figure2}}
{\includegraphics[width=0.5\textwidth]{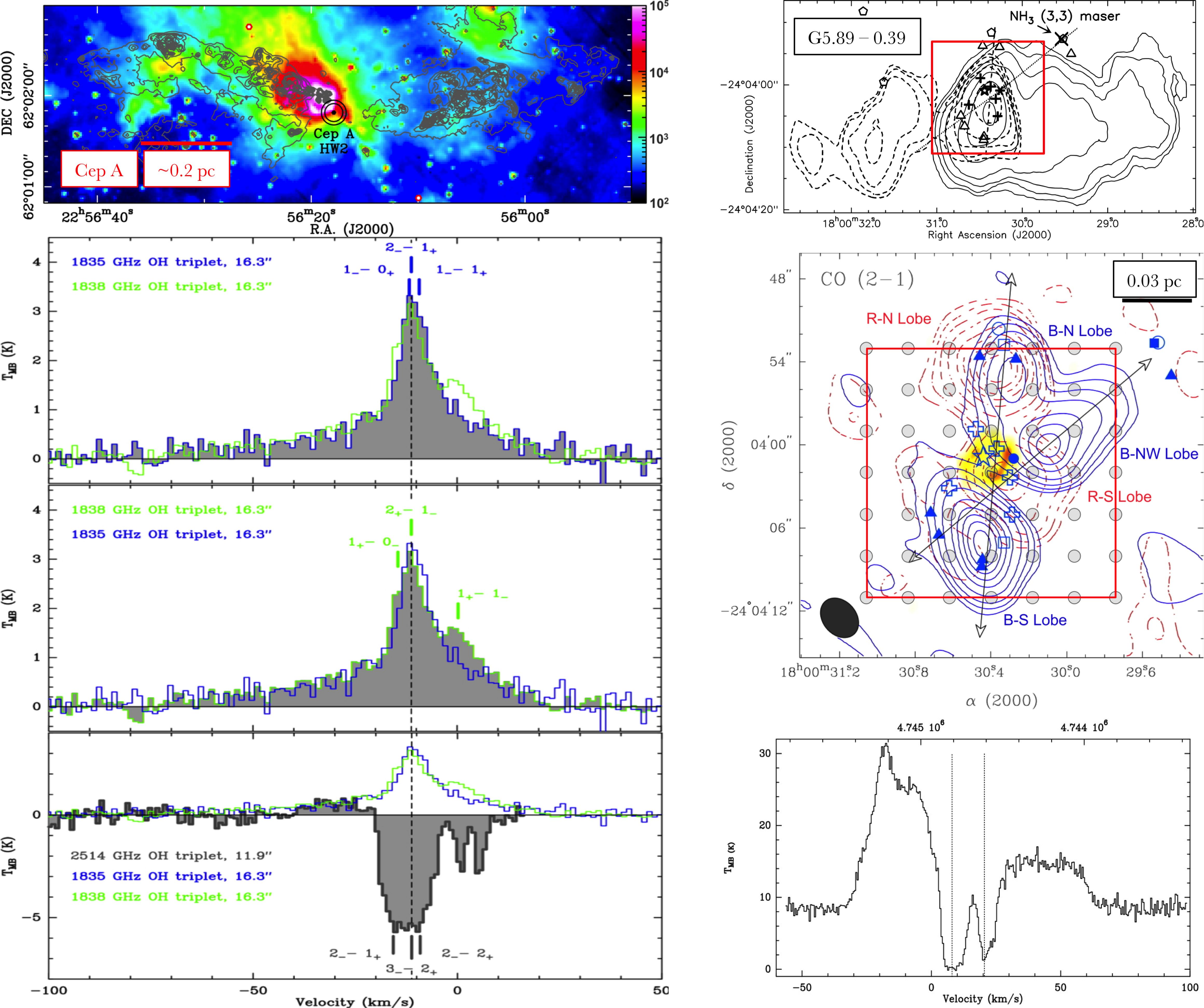}}
\end{figure}

When climbing the protostellar mass ladder, new questions are asked. The scenario for massive star formation is less clear (is it a scaled-up version from the low-mass one ? Do massive stars result from competitive accretion mechanisms, or should other mechanisms be invoked ?), and the role of the important protostellar radiation field must be quantified. To these aims, observing CO is still paramount, as well as its photodissociation and photoionization products: C and C$^+$. This is also true for a molecule widely observed by \textit{Herschel}, H$_2$O: OH and OI. High-frequency, velocity-resolved observations of such tracers (C$^+$, OH, and OI) can only be performed with SOFIA/GREAT. In the Cep E intermediate-mass jet/outflow system (see Fig.~\ref{figure1}, Gomez-Ruiz {\em et al.\/} \cite{Gomezruiz12} and Lefloch {\em et al.\/} \cite{Lefloch15}), CO observations by SOFIA/GREAT were used to show that the influence of the protostellar radiation field is negligible, at least in the terminal shock regions. In the Cep A and G5.89--0.39 (Gusdorf {\em et al.\/} \cite{Gusdorf152}, Gusdorf {\em et al.\/} \cite{Gusdorf153}, Leurini {\em et al.\/} \cite{Leurini15}) massive star-forming regions, OH, OI, and C$^+$ observations were used to perform detailed energetics evaluations, to highlight the shortcomings of current shock models and to illustrate the necessity of developing \lq irradiated' shock models. In a forthcoming steps, these observations will serve as a basis for more thorough chemical studies, specially on the chemistry of H$_2$O.

\section{The \textit{Herschel} legacy: filaments and ridges}
\label{sec:thlfar}

\begin{figure}
\centering
\includegraphics[width=0.85\textwidth]{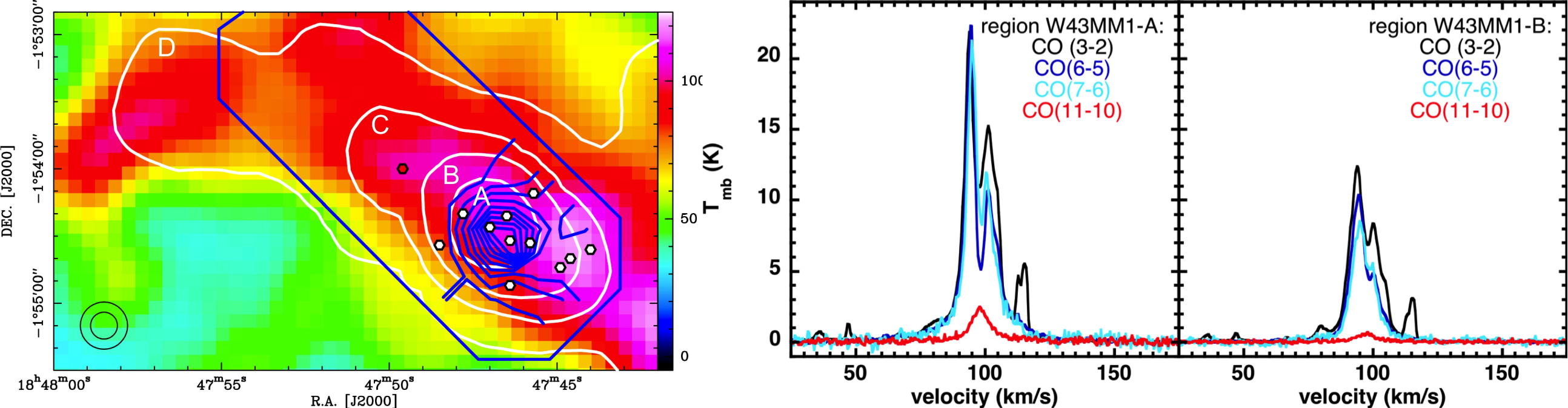}
\caption{\textit{Left-hand panel:} the W43-MM1 ridge: integrated intensity in $^{13}$CO (2-1) (Carlhoff {\em et al.\/} \cite{Carlhoff13}, colour map) and $^{12}$CO (11-10) (Gusdorf {\em et al.\/}, in prep., blue contours). The white contours delimitate the regions A, B, C, and D defined in Louvet {\em et al.\/} \cite{Louvet14} based on column density values inferred from \textit{Herschel} observations. The white hexagons mark the massive dense cores identified by Louvet {\em et al.\/} \cite{Louvet14} based on continuum observations from the PdBI. The red hexagon is a dense core. \textit{Right-hand panel:} $^{12}$CO spectra over a 24'' beam size, extracted from regions A and B (Gusdorf {\em et al.\/}, in prep.).}
\label{figure3}
\end{figure}

A significant legacy of \textit{Herschel} is the study of filaments and ridges. These structures are the result of dynamical processes of matter accretion, themselves at the base of the formation of molecular clouds: colliding flows. The ridges are the densest filaments, where clustered, massive star formation is usually observed (see Fig.~\ref{figure3} and e.g. Nguyen-luong {\em et al.\/} \cite{Nguyenluong13}). Their observation yields numerous questions: what is the chemistry that takes place in such regions of low-velocity, dense shocks ? What are the accretion mechanisms leading to the formation of the very massive protostars that seem to be detected ? Can the most sophisticated models of ISM evolution account for the formation of such filaments with embedded massive stars ? What is the contribution of these low-energy but extended regions to the high-$J$ CO emission as seen by \textit{Herschel} on a larger, galactic scale~? Again, only SOFIA/GREAT can provide the answers to such questions. In Louvet {\em et al.\/} \cite{Louvet14}, we used PdBI observations to compare ongoing star formation with large simulations. In Gusdorf {\em et al.\/}, we will use CO and C$^+$ observations by SOFIA/GREAT to characterize the shocks, and directly evaluate their contribution to the galactic emission. These results will support our new models of SiO formation in low-velocity shock environments (Louvet {\em et al.\/}, in prep.).


\end{document}